\documentclass[conference]{IEEEtran}

\usepackage{hyperref}
\hypersetup{
     colorlinks = true,
     citecolor = green,
     }

\ifCLASSINFOpdf
  \usepackage[pdftex]{graphicx}
\else
\fi

\usepackage{framed}
\begin{document}
%
\title{Digital Twins Are Not Monozygotic -- Cross-Replicating ADAS Testing in Two Industry-Grade Automotive Simulators}



%
\author{\IEEEauthorblockN{Markus Borg\IEEEauthorrefmark{1},
Raja Ben Abdessalem\IEEEauthorrefmark{2},
Shiva Nejati\IEEEauthorrefmark{3}\IEEEauthorrefmark{2}, 
Fran\c cois-Xavier Jegeden\IEEEauthorrefmark{4} and
Donghwan Shin\IEEEauthorrefmark{2}}
\IEEEauthorblockA{\IEEEauthorrefmark{1}RISE Research Institutes of Sweden and Lund University, Lund, Sweden, markus.borg@ri.se}
\IEEEauthorblockA{\IEEEauthorrefmark{2}University of Luxembourg, Luxembourg, Luxembourg. raja.benabdessalem@uni.lu, donghwan.shin@uni.lu}
\IEEEauthorblockA{\IEEEauthorrefmark{3}University of Ottawa, Ottawa, Canada, 
snejati@uottawa.ca}
\IEEEauthorblockA{\IEEEauthorrefmark{4}ESI Group, Nantes, France, Francois-Xavier.Jegaden@esi-group.com}}


\maketitle

\begin{abstract}
The increasing levels of software- and data-intensive driving automation call for an evolution of automotive software testing. As a recommended practice of the Verification and Validation (V\&V) process of ISO/PAS 21448, a candidate standard for safety of the intended functionality for road vehicles, simulation-based testing has the potential to reduce both risks and costs. There is a growing body of research on devising test automation techniques using simulators for Advanced Driver-Assistance Systems (ADAS). However, how similar are the results if the same test scenarios are executed in different simulators? We conduct a replication study of applying a Search-Based Software Testing (SBST) solution to a real-world ADAS (PeVi, a pedestrian vision detection system) using two different commercial simulators, namely, TASS/Siemens PreScan and ESI Pro-SiVIC. Based on a minimalistic scene, we compare critical test scenarios generated using our SBST solution in these two simulators. We show that SBST can be used to effectively generate critical test scenarios in both simulators, and the test results obtained from the two simulators can reveal several weaknesses of the ADAS under test. However, executing the same test scenarios in the two simulators leads to notable differences in the details of the test outputs, in particular, related to (1) safety violations revealed by tests, and (2) dynamics of cars and pedestrians. Based on our findings, we recommend future V\&V plans to include multiple simulators to support robust simulation-based testing and to base test objectives on measures that are less dependant on the internals of the simulators.
\end{abstract}


%

\section{Introduction}
There is a growing trend to increase the level of vehicle automation driven by the recent advances in technologies such as, among others, Machine Learning (ML) and Deep Neural Networks (DNN), computer vision, and sensor fusion. However, in parallel with this technological growth, there is an increase in the number of accidents and crashes that involve self-driving cars and pedestrians~\cite{selfdriving}. Many of these accidents are due to an interplay between software, often containing complex ML-based components, and advanced electronics, e.g., cameras and LiDAR technologies, that are used in today's modern vehicles. To prevent such accidents and crashes, there is a need to perform Verification and Validation (V\&V) techniques for self-driving vehicles at \emph{system-level} to ensure that they are safe and reliable before letting them drive on public roads~\cite{borg2019safely}. 

Currently, in industry, the major bulk of system-level testing of self-driving vehicles is carried out through on-road testing or using naturalistic field operational tests. These activities, however, are expensive, dangerous, and ineffective~\cite{koopman:16}. A feasible and efficient complementary approach is to conduct system-level testing through computer simulations that can capture the entire self-driving vehicles and their operational environment using effective and high-fidelity physics-based simulators. There is a growing number of commercial and public-domain simulators that have been developed over the past few years to support realistic simulation of self-driving systems~\cite{math2013opends,dosovitskiy2017carla,shah2018airsim}. In the
ISO/PAS 21448 Safety of the Intended Function (SOTIF) candidate standard~\cite{international_organization_for_standardization_road_2019}, an ongoing standardization initiative covering automotive ML, simulation is recognized as one of the main V\&V means for self-driving cars. This has led to the development of a large number of system-level testing approaches in the literature that rely on such simulators. 

Existing testing techniques are often focused on devising algorithms and techniques to generate test cases~\cite{gambi_generating_2019, ben_abdessalem_testing_2018-1, ben_abdessalem_testing_2016} or to generate test oracles~\cite{Stocco20}. There is, however, little research on studying the role of simulators when testing is based on a simulation environment. Recently, Sotiropoulos~\textit{et al}.~\cite{SotiropoulosWGI17} provided an empirical study comparing testing results of robot function models obtained based on a simulator with those obtained from their physical field testing. Ul~Haq~\textit{et al.}~\cite{HaqSNB20} and Codevilla~\textit{et al.}~\cite{Codevilla_2018_ECCV} compare testing of DNN-based automated driving systems based on real-world and simulator-generated images and videos. We pose a cornerstone question that has not been previously studied in the simulation-based testing literature: Can we obtain similar or consistent test results from different simulators? Answering this question requires replicating testing techniques in different simulators and studying the results. We refer to such studies as  \emph{cross-simulator (X-sim)} replications.

Ben Abdessalem \textit{et al.} have conducted several studies on ADAS testing using the simulator TASS/Siemens PreScan~\cite{ben_abdessalem_testing_2016, ben_abdessalem_testing_2018, ben_abdessalem_testing_2018-1}. These papers show how Search-Based Software Testing (SBST)~\cite{mcminn2011search} can be used to effectively  find input values to generate test scenarios that stress individual ADAS components~\cite{ben_abdessalem_testing_2016}. The stress test scenarios, which are also referred to as critical test scenarios, are obtained such that they break or are close to breaking safety requirements of the ADAS under test, and hence, result in a \textit{safety violation}.

In this paper, we investigate if the results obtained from ADAS testing are consistent across different simulators. 
To this end, we present a X-sim replication study in which we ported the solution by Ben Abdessalem \textit{et al.}~\cite{ben_abdessalem_testing_2016} to the ESI Pro-SiVIC simulator~\cite{belbachir_simulation-driven_2012} which is an alternative commercial automotive simulator. The \textit{original study} applies SBST to an ADAS example, i.e., the Pedestrian Detection Vision based system (PeVi). Specifically, the original study is focused on testing PeVi using simulations capturing the ego car (i.e., the car augmented with ADAS) driving on a straight urban street and a pedestrian crossing the street from the right. We adhere to the definitions of \emph{scene} and \emph{scenario} proposed by Ulrich \textit{et al.}~\cite{ulbrich_defining_2015}, which are also used in SOTIF. A scene is ``a snapshot of the environment including the scenery, dynamic elements, and all actor and observer self-representations, and the relationships between those entities''. A scenario describes ``the temporal development between several scenes in a sequence of scenes''. In collaboration with the original authors, we simplified the PreScan scene that was used for test generation to support porting to Pro-SiVIC with minimal differences. By controlling as many variables as possible related to the (initial) scene, we focus this study to compare the scenarios generated based on the initial scene. In addition, we ported PeVi to Pro-SiVIC so that the replication and the original study use the same ADAS under test. 

In line with the terminology used by Cartwright~\cite{cartwright_replicability_1991} and Gomez \textit{et al.}~\cite{gomez_replication_2010}, we refer to our work as a series of \textit{reproductions}. Three research questions guide our study:
\begin{itemize}
\item[RQ1] Is SBST an effective approach to ADAS testing if we replace PreScan  with Pro-SiVIC?
\item[RQ2] Is the diagnostic information obtained by applying SBST using PreScan reproducible if we use Pro-SiVIC? 
\item[RQ3] Given a minimalistic scene, to what extent can critical test scenarios identified in PreScan be reproduced in Pro-SiVIC, and vice versa?
\end{itemize}

Our results show that SBST can be used to effectively generate critical test scenarios in both simulators, and the test results obtained from the two simulators can reveal several weaknesses of PeVi (the ADAS under test). However, the test scenarios obtained by PreScan and Pro-SiVIC do not lead to consistent and conclusive characterizations of safety violations for PeVi. In particular, the only consistent diagnostic information that we identify in our study is that, in both PreScan and Pro-SiVIC, PeVi likely violates its  safety requirement when the car moves fast (more than 72 km/h). Finally, reproducing critical scenarios between PreScan and Pro-SiVIC can result in discrepancies that might not only be due to the implementation of PeVi, but can originate in differences in the dynamic models of the simulators or the off-the-shelf sensors available in the simulators’ libraries. This research concludes by two lessons-learned and recommendations that have the potential to influence future simulation-based testing of ADAS. 

\emph{Paper organization.} Section~\ref{sec:bg} presents related work on ADAS testing and introduces the original study. The process of porting scenes to Pro-SiVIC is described in Section~\ref{sec:porting}. Section~\ref{sec:method} explains the research method and Section~\ref{sec:res} presents the results. Finally, Section~\ref{sec:threats} discusses the main threats to validity, Section~\ref{sec:lessons} provides lessons learned, and Section~\ref{sec:conc} concludes the paper.

\section{Background and Related Work} \label{sec:bg}
This section presents a brief overview of related work and details about the original study reproduced in this paper.

\subsection{Simulation-based CPS and ADAS Testing}
Digital twins~\cite{8972429} are defined as digital and virtual representations of physical assets enabled through data and simulators for monitoring, controlling, optimization and verification purposes. There is an increasing demand for fast, agile and high fidelity digital twins in the domain of cyber physical systems (CPS). For ADAS and self-driving systems, there is even a higher demand for digital twins and simulators since real-world testing and verification of such systems is expensive, dangerous, and ineffective. 
Various simulators such as those relying on physics-based modeling (e.g., Pro-SiVIC and PreScan) or those that rely on game engines (e.g., ~\cite{gambi_generating_2019,gambi_automatically_2019}) have been used for testing of self-driving systems and ADAS. Due to the large search space for ADAS and self-driving systems, achieving any form of coverage over the space of all possible simulation scenarios is rather infeasible. Hence, search-based software testing has been advocated as an effective and efficient strategy to generate test scenarios for such systems when they are tested within a simulation environment~\cite{gambi_generating_2019,gambi_automatically_2019,ben_abdessalem_testing_2016,buhler:04}. While the focus of the current research is on devising testing techniques, in this paper, we evaluate the impact of simulators on the test results through a replication study performed using two physics-based ADAS simulators: PreScan and Pro-SiVIC. 

\subsection{Description and Definitions of the Original Study}\label{subsec:original}

The original study used SBST and PreScan to test PeVi as part of an industrial ADAS case study~\cite{ben_abdessalem_testing_2016}. In this section, we provide details of the original work. 

\begin{figure}
\centering
\includegraphics[width=0.45\textwidth]{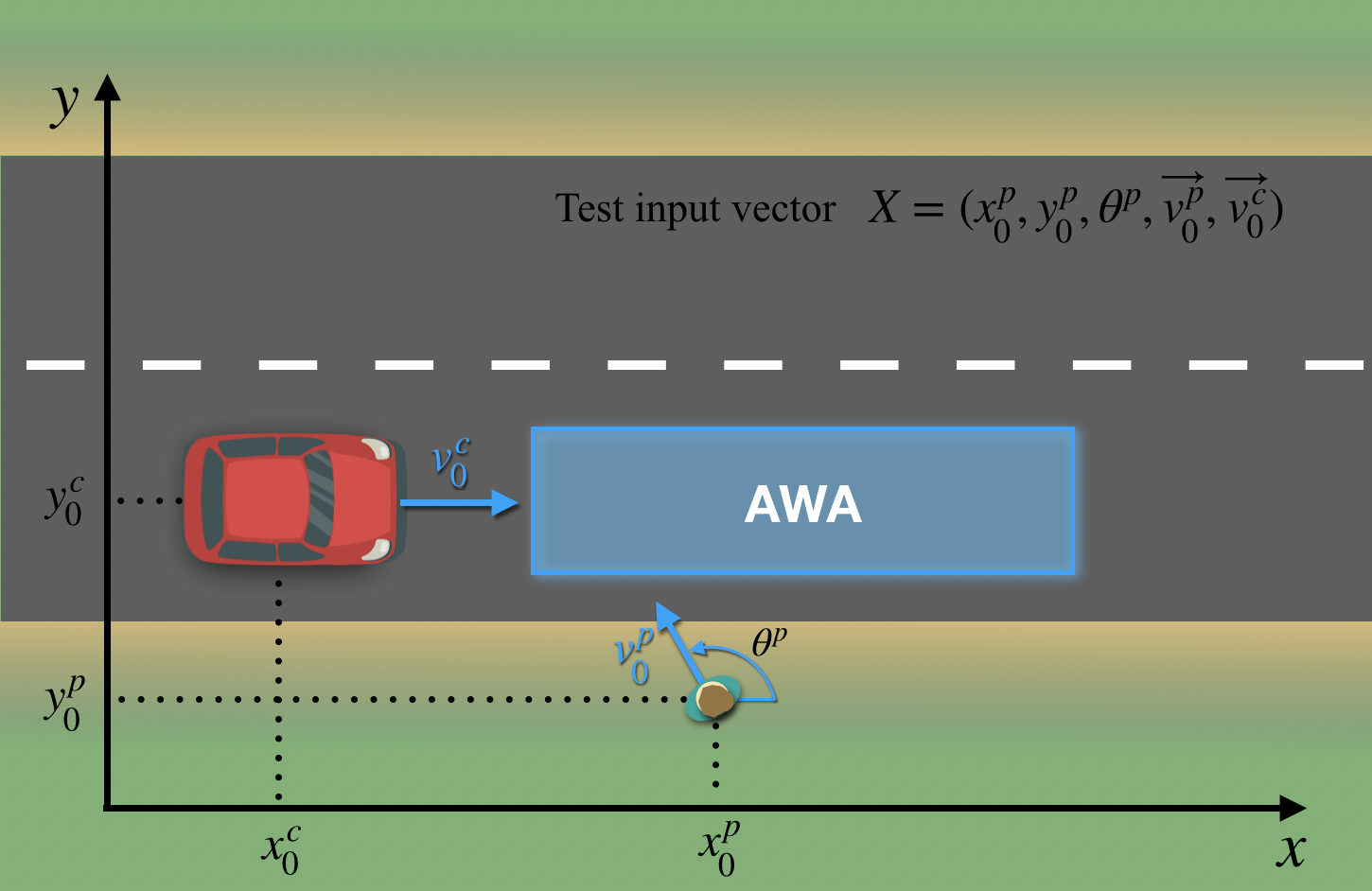}
\caption{Input variables in the initial scene.}
\label{fig:input}
\end{figure}

\textbf{Study subject.} Briefly, PeVi's function is to determine whether there is any pedestrian in a rectangular \emph{Acute Warning Area} (AWA) in front of the car, and if so, it shows a warning message to the driver. The size of the AWA depends on the speed of the car and the shape of the road. Figure~\ref{fig:input} shows the AWA for a car driving on a straight road. PeVi uses data received from a sensor component to identify the position and the speed of the objects in or near the AWA. It also receives the Time To Collision (TTC) as computed by the sensor component. TTC measures the time until impact between the ego car and an object if both continue with the same velocities~\cite{van:93}. When an object is detected in or near AWA ($\leq 0.2 m$ from the boundaries), and when the TTC is below a defined threshold, the object position is sent to the camera to detect object types and shapes after receiving their positions from the sensor component. Specifically, the vision component determines whether the object is a pedestrian. Then, PeVi will show a warning message to the driver indicating that the car may risk a collision with a pedestrian. 

\textbf{Safety Requirement.} The test generation aims to verify the following functional safety requirement of PeVi: \emph{``PeVi shall detect pedestrians in or near AWA ($\leq 0.2 m$ from the boundaries) when there is a risk of collision with the pedestrians and when the pedestrians are close to the car''}. 

This requirement originated with customers (car manufacturers) where the statement ``there is a risk of collision with the pedestrians and when the pedestrians are close to the car" was not detailed. 
As we describe later, the above requirement is detailed and formalized using quantifiable fitness functions through interactions with the engineers who developed PeVi. 

\textbf{Scope of Testing.} A number of simulation-based testing studies~\cite{ben_abdessalem_testing_2018-1,ben_abdessalem_testing_2018}, vary both static elements (e.g., different weather conditions, different road shapes, and different background scenes) and  dynamic elements of simulators (e.g., the speed, the position, and the trajectory of cars and pedestrians). The original study, however, fixed the initial scene to include the ego car driving on a straight urban street and a pedestrian crossing the street from the right. The test generation then focuses on varying the dynamics, namely, the speed of the car, and the speed, position and orientation of the pedestrian. Replicating this study (using an even further simplified scene) allows us to focus on comparing the generated scenarios within a plain and simple scene.

Note that we have to include PeVi in a simulation environment and perform system-level testing to verify PeVi against its safety requirement. However, the faults identified using system-level testing may not necessarily be due to faults or errors in PeVi's implementation and may be due to errors in the simulators or in third party models of hardware components (e.g., sensors and cameras), or due to the real world and physical constraints~\cite{ben_abdessalem_testing_2018}. 

\textbf{Input Representation.} According to the original study, the test input space of PeVi, which is also depicted in Figure~\ref{fig:input}, consists of vectors ($v^c_0, x_0^p, y_0^p, \theta^p, v_0^p$) where $v_0^c$ is the car speed, $x_0^p$ and $y_0^p$ specify the (initial) position of the pedestrian, $\theta^p$ is the orientation of the pedestrian, and $v_0^p$ is the pedestrian speed.  Note that the initial car position is fixed at $(x^c_0, y^c_0)$. The variables in the search space are further constrained as follows: $1 \leq v_0^c~(m/s) \leq 25$;  $x_0^c+20 \leq x_0^p~(m) \leq x_0^c+85$; $y_0^c-15 \geq y_0^p~(m) \geq y_0^c-2$; $40 \leq \theta^p(^\circ) \leq 160$ and $1 \leq v_0^p~(m/s) \leq 5$.  Each value assignment to the vector ($v_c, x_0, y_0, \theta,v_p$) represents a test input for PeVi.

\textbf{Fitness Functions}. SBST exercises PeVi with respect to its requirement guided by minimizing three fitness functions:
\begin{itemize}
    \item[FF1] The minimum distance between the ego car and the pedestrian over the test scenario.
    \item[FF2] The minimum distance between the AWA and the pedestrian over the test scenario.
    \item[FF3] The minimum TTC between the ego car and the pedestrian over the test scenario. 
\end{itemize}
The outputs of each test scenario (simulation) include a vector of distances at each simulation time step between: the ego car and the pedestrian and the AWA and the pedestrian, as well as a vector of TTCs at each simulation time step between the ego car and the pedestrian. We select the minimum value of these vectors to compute the fitness functions. Note that test scenarios do not stop upon detection by PeVi. We run each test scenario for a time duration and stop them when any of these conditions holds: (1) the car has driven 100 m (i.e., the length of the road segment under analysis), or (2) the pedestrian has crossed the road, or (3) the car has passed the pedestrian (either there was a collision or the pedestrian did not yet reach the road). The original study discusses how \emph{minimizing} the above three fitness functions pushes PeVi into breaking its requirement. 

\textbf{The Computational Search Algorithm.} The search algorithm used to test PeVi is the Non-dominated Sorting Genetic Algorithm version 2 (NSGA-II)~\cite{deb1995simulated}, a well-known multi-objective search algorithm that has been used in many different domains. Note that we need to use a multi-objective search algorithm to test PeVi, since breaking the safety requirement of PeVi requires us to  minimize the three fitness functions defined above. The following summarizes the choice of operations and parameters of NSGA-II used in the original study~\cite{ben_abdessalem_testing_2016}:

\begin{itemize}
    \item Selection. We use a binary tournament selection with replacement that has been used in the original implementation of the NSGA-II algorithm.
    \item Crossover. We use the Simulated Binary Crossover operator (SBX). SBX creates two offsprings from two selected parent individuals. The difference between offsprings and parents is controlled by a distribution index ($\eta$): When $\eta$ is large, the offsprings are closer to the parents, while a small $\eta$ increases the difference. Analagous to the original study, we chose a high value for $\eta$ (i.e., $\eta = 20$) based on the guidelines by Deb and Agrawal~\cite{deb1995simulated}.
    \item Mutation. Mutation is applied after crossover to the genes of the children chromosomes with a certain probability (mutation rate). Given a gene $x$ (i.e., any of the variables $v^c_0$, $x_0^p$, $y_0^p$, $\theta^p$, $v_0^p$), our mutation operator shifts $x$ by a value selected from a normal distribution with mean $\mu = 0$ and variance $\sigma^2$. To avoid invalid offsprings from crossover or mutation, we use cutoffs corresponding to the end points of the ranges.
\end{itemize}

The NSGA-II search parameters were selected as follows: the crossover rate was set to 0.9, the mutation rate to 0.5, and the population size to 10. In this replication study, we reuse the same search parameters as in the original study.

\textbf{Testing Time Budget.} In the original study, NSGA-II ran within a restricted execution time budget of 150 min. The time budget was selected in consultation with the   supplier of PeVi. The experiments reported in the original study show that the time budget was sufficient to find failure revealing test scenarios and also to demonstrate that NSGA-II outperforms random search testing (the sanity check experiment in SBST~\cite{Harman:12}). We use the same time budget for the replication study.

\textbf{Critical and Safety Violation Scenarios.}
In this replication study, we discuss critical scenarios and safety violations using the following definitions. A \emph{critical} scenario either results in a collision between the ego car and the pedestrian or a near miss. A near miss is defined as a scenario with FF1 $\leq$ 1 m or FF3 $\leq$ 0.5 s. In addition, we define a \emph{safety violation} as a \emph{critical} scenario where PeVi has failed to detect the pedestrian. Note that in critical scenarios, PeVi may or may not have detected the pedestrian. We are interested in generating both critical and safety violation scenarios. While safety violation scenarios indicate clear violations of the PeVi requirement, critical scenarios represent situations where the car and pedestrian may have a collision or a near miss. For a critical scenario, even if there is a detection, it is important to know the time gap between the detection and collision (or near miss) to determine if the gap is sufficiently large so that the driver can react and avoid the collision. 

\section{Porting from PreScan to Pro-SiVIC} \label{sec:porting}
Together with the lead developer of the original study~\cite{ben_abdessalem_testing_2016}, we ported the SBST algorithm as well as the PeVi component from PreScan to Pro-SiVIC. The process required considerable engineering effort, made possible through physical co-location during a two month research visit. This section describes similarities and differences between the two simulation setups.

\begin{figure}
\centering
\includegraphics[width=0.475\textwidth]{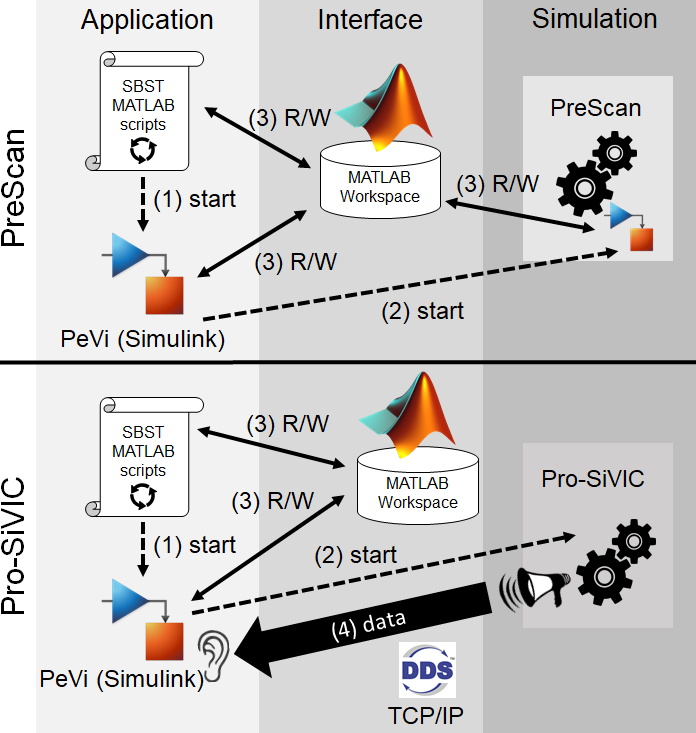}
\caption{Simulation setups used for PreScan and Pro-SiVIC. Dashed arrows depict function calls that happen once per scenario, solid arrow show read/write operations to shared variables, and the big arrow represents DDS communication.}
\label{fig:setups}
\end{figure}

Figure~\ref{fig:setups} depicts the simulation setups used for ADAS testing with PreScan and Pro-SiVIC, respectively. The figure is organized into three layers:
\begin{itemize}
    \item \textbf{Application} -- MATLAB/Simulink implementations of the SBST algorithm and PeVi.
    \item \textbf{Interface} -- The interface between the simulator and both the SBST algorithm and PeVi. 
    \item \textbf{Simulation} -- The simulator tool which includes mechanisms to construct initial scenes as well as physics-based and mathematical models that simulate sensors and dynamic objects such as cars and pedestrians. 
\end{itemize}

For the \textit{Application} layer, porting the original implementation of the SBST algorithm was straightforward. For both PreScan and Pro-SiVIC, MATLAB scripts implement the NSGA-II algorithm and call the Simulink model of PeVi to initialize it with specific test inputs once per test scenario. The PeVi model, in turn, calls the simulator to start  generating the output corresponding to the given test input (see links labelled (1) and (2) in Figure~\ref{fig:setups}). 
In the Application layer, there is a one-to-one mapping between the elements used in the PreScan setting and those used in the Pro-SiVIC setting. PeVi from the original study was reused without modifications in the replication study. Still some engineering work was needed, primarily in relation to configuring the sensor model of Pro-SiVIC and some data type conversion to ensure that Pro-SiVIC could generate the input formats required by PeVi.

The main differences between the PreScan and Pro-SiVIC setups are related to the \textit{Interface} layer. PreScan uses Simulink internally for both modeling the physics and the motion behavior of vehicles and pedestrians as well as sensor modeling -- a local Simulink installation is even a prerequisite to run PreScan. As a result, external Simulink models (such as the model of PeVi) can easily be integrated with PreScan since they can read and write to shared variables in the same MATLAB workspace (see the links labelled (3) in Figure~\ref{fig:setups}). 

In contrast, Pro-SiVIC does not depend on Simulink for internal modeling. In Pro-SiVIC, elements communicate through the Data Distribution Service (DDS)~\cite{dds2015}, a message-based middleware protocol implementing a publish-subscribe pattern (see the link labelled (4) in Figure~\ref{fig:setups}). Hence, PeVi communicates with the internal models of PreScan synchronously, while the DDS-based communication between PeVi and Pro-SiVIC is asynchronous. To initiate the communication, the external Simulink model (PeVi) starts the Pro-SiVIC scenario, and then, each DDS enabled element in Pro-SiVIC (e.g., the sensors and the car) begins broadcasting DDS messages to the subscribing Simulink blocks of PeVi each 40 ms (25 Hz).

The Simulink simulation was not fast enough to receive DDS messages at this frequency. We measured a uniform packet loss of 20\%, i.e., roughly every fifth DDS message was not received by PeVi. This means that 20\% of the DDS messages corresponded to an 80 ms measurement interval instead of 40 ms. To mitigate the risk of losing DDS messages containing data with a minimum FF, each Pro-SiVIC scenario was repeated 20 times and the mode of the FF measurements (i.e., the most frequently generated outputs) were used for the subsequent analysis. We found that this mitigation strategy generated simulation results that were not impacted by the packet loss. Indeed, in these scenarios, due to the high frequency of the messages sent from Pro-SiVIC, the content of the lost messages were redundant or were very similar to the messages coming immediately before or after them and processed by the Simulink model. 



For the \textit{Simulation} layer, there are inevitable differences between the initial scenes in PreScan and Pro-SiVIC. As we discussed in Section~\ref{subsec:original}, one of the reasons that we choose the study of Ben Abdessalem~\cite{ben_abdessalem_testing_2016} for replication is because this study is mainly focused on varying simulation dynamics (i.e., the position and speed of objects) and the background scene is unchanged over different test scenarios.  
To mitigate the potential threats to internal validity and to make sure that we compare the dynamic behavior of simulators rather than their motifs and initial scene construction abilities, we reduced the complexity of the  scene of PreScan used in the original study~\cite{ben_abdessalem_testing_2016}, i.e., we created a novel \emph{minimalistic} PreScan scene with removed buildings along the road and no shadows from the pedestrian. In PreScan, we built the minimalistic scene from scratch. On the other hand, Pro-SiVIC scenes are typically built using existing road snippets or adapted from pre-made standard scenes. Thus, we implemented the minimalistic scene using a straight road segment from the standard scene ``horsering-ground'' with a similar skydome and illumination settings as in the PreScan scene.

\begin{figure}
\centering
\includegraphics[width=0.45\textwidth]{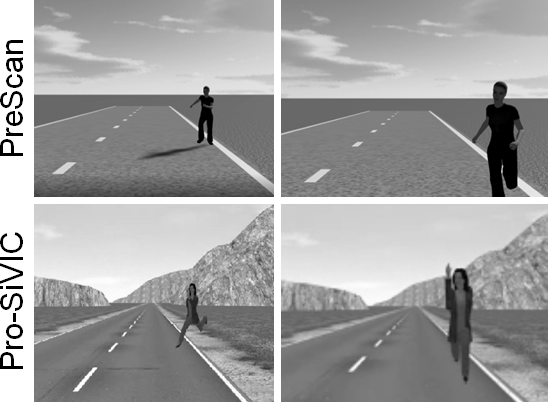}
\caption{Crossing pedestrians in PreScan and Pro-SiVIC. Note that we later disabled shadows in PreScan.}
\label{fig:compare}
\end{figure}

While we attempted to create equivalent initial scenes, some differences are obvious, including the visual appearance of approaching pedestrians as shown in Figure~\ref{fig:compare}. We report three major differences:
\begin{enumerate}
    \item The default pedestrian in PreScan is male, whereas the Pro-SiVIC pedestrian is female. The man runs with a swinging arm movement while the woman pumps the arms like a sprinter. Furthermore, the pedestrians wear different clothes.
    \item The horizon is visible in the PreScan scene, while in Pro-SiVIC, it is occluded by mountains in the distance.
    \item The road in the Pro-SiVIC scene has a narrow dirt shoulder, but the PreScan scene has no shoulder at all.
\end{enumerate}

Finally, we used the same test input characterization with the same constraints and ranges when generating test scenarios for both PreScan and Pro-SiVIC. However, as the coordinates for the initial position of the car differ between the two scenes in PreScan and Pro-SiVIC, we implemented a translation function. Furthermore, similar translations were needed for the orientation of the pedestrian and conversions between m/s and km/h. All source code for the replication study is available on GitHub under a BSD 2-Clause~\cite{github2020}. The source code and data related to the original study is available on BitBucket~\cite{bitbucket2020}.

\section{Research Method} \label{sec:method}
This section describes the design of the empirical study.

\subsection{Experimental Design}
\begin{figure}
\centering
\includegraphics[width=0.8\linewidth]{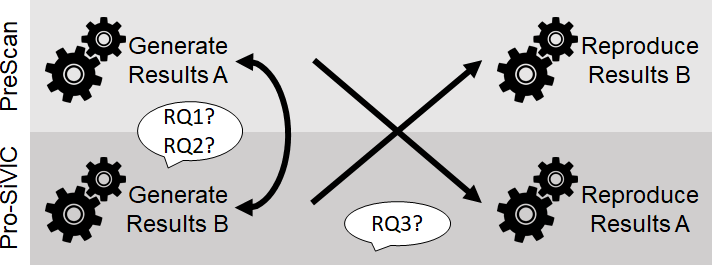}
\caption{Comparisons relevant to the RQs.}
\label{fig:rqs}
\end{figure}

Figure~\ref{fig:rqs} shows an overview of our experimental process. While we describe our empirical work as sequential steps, most experiments with PreScan and Pro-SiVIC were conducted in parallel -- typically running overnight due to long execution times.

\subsubsection{RQ1 -- X-sim Reproduction of Principal Findings}
RQ1 concerns the high-level replicability of the original study~\cite{ben_abdessalem_testing_2016}. Can we show that SBST enabled by NSGA-II is an effective approach to ADAS testing even if we replace PreScan with another simulator? 

To answer this question, we executed both the PreScan and the Pro-SiVIC setups for 40 times to account for the randomness in the NSGA-II algorithm. For each run of each setup, we used the testing time budget of 150 min from the original study. Note that each run of NSGA-II, being a multi-objective algorithm, generates 10 solutions (i.e., equivalent to the population size for NSGA-II provided in Section~\ref{subsec:original}). Thus, we obtained 400 scenarios in total.

To answer RQ1, we analyze the outputs from the PreScan and Pro-SiVIC setups to compare the quality of the generated test cases. In particular, we want to determine if SBST can generate fault revealing and critical test cases in both setups. We use two types of metrics for this purpose: (1) The number of test scenarios representing a critical or safety violation situation (see Section~\ref{subsec:original} for the definitions of critical and safety violations). (2) An assessment of the NSGA-II outputs using the hypervolume (HV) indicator~\cite{Brockhoff:08}. The HV indicator has been commonly used in the literature (including the original study~\cite{ben_abdessalem_testing_2016}) to evaluate multi-objective search algorithms since their outputs create a Pareto front~\cite{deb1995simulated}. 
Briefly, HV represents the size of the space covered by members of a Pareto front generated by a search algorithm~\cite{Brockhoff:08}. The higher the HV values, the better the Pareto front outputs are. To compare the statistical differences in HV values generated by PreScan and Pro-SiVIC, we use the Mann–Whitney U test at $\alpha=0.05$. 


\subsubsection{RQ2 -- X-sim Reproduction of Diagnosis Information} While RQ1 is focused on the reproduction of test outputs in the two simulators, RQ2 investigates the consistency of the diagnostic information that can be derived from the test inputs generated by the application of SBST in the two simulators. In general, there is little research on producing diagnosis or debugging support for self-driving systems and ADAS. One proposed approach is to apply classification decision trees to identify conditions on test inputs that best explain and characterize failures~\cite{ben_abdessalem_testing_2018}. Decision tree learning is a supervised learning classification technique~\cite{witten:11}. To answer RQ2, we use the same results generated by the experiment we performed for RQ1. But this time, we study the distributions of the test inputs, and further, we use a decision tree classifier to infer conditions on the test inputs that can best characterize safety violations in the two simulators. 

\subsubsection{RQ3 -- X-Sim Reproduction of Critical Test Scenarios} \label{sec:designrq2}
RQ3 addresses the reproduction of test scenarios in another simulator. If a scenario is found to be critical in PreScan, will the same scenario also be critical in Pro-SiVIC and vice versa? Recall that Section~\ref{subsec:original} provides the definitions for critical scenarios and safety violations. Our goal is to understand to what extent test inputs leading to critical scenarios or safety violations in one simulator remain critical or yield safety violations when executed in another simulator. 

For this question, we converted the test inputs corresponding to the 400 test scenarios generated in PreScan for RQ1 to their Pro-SiVIC counterparts (as described under the Application layer in Section~\ref{sec:porting}). The converted test scenarios were then executed in Pro-SiVIC. Then, we repeated the analogous procedure to re-execute the scenarios generated by  Pro-SiVIC in PreScan.


To answer RQ3, we analyze the outputs from PreScan and Pro-SiVIC from two perspectives: (1) the fraction of safety violations that remain after X-sim reproduction and if any new appear, and (2) the absolute differences of the results from the three fitness functions (FF1, FF2, and FF3) when reproducing scenarios across simulators.


\subsection{Hardware and Software Setups}
The PreScan and Pro-SiVIC setups used standalone licenses linked to the physical MAC addresses of specific devices. As the simulator vendors granted licenses to different organizations, we were not able to install a license server accessible over an internal network to execute PreScan and Pro-SiVIC on the same device. As a result, we conducted the simulations on separate computers. While this might introduce confounding factors, we believe this does not have an impact on our conclusions since our analysis is not focused on computational performance. In particular, in our analysis, we do not compare the time performance of the two simulator setups. The setup used to run the PreScan experiments was a MacBook Pro with a 2.5 GHz CPU and 16 GB RAM with PreScan version 2019.1 and MS Windows 10. We conducted the Pro-SiVIC experiments on a desktop PC running MS Windows 10 equipped with an Intel Core i7-3770 CPU @ 3.40 GHz, 32 GB RAM, and an Nvidia 1080Ti graphics card. The software version used was ESI Pro-SiVIC 2018.0.

\section{Results and Discussion} \label{sec:res}
This section presents results from the X-sim  reproductions and discuss their practical implications.

\subsection{RQ1: X-sim Reproduction of Principal Findings} \label{sec:resrq1}
\begin{figure}
\centering
\includegraphics[width=0.48\textwidth]{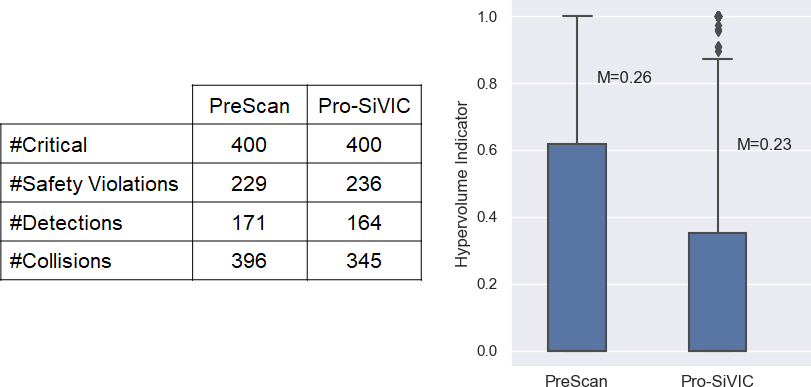}
\caption{Overview of the 800 generated test scenarios by PreScan and Pro-SiVIC. The boxplots show the hypervolume indicators, reflecting the quality of the Pareto fronts.}
\label{fig:table_hv}
\end{figure}

Figure~\ref{fig:table_hv} (the left part) presents the number of critical scenarios and safety violations generated in PreScan and Pro-SiVIC (see Section~\ref{subsec:original} for the definitions of safety violation and critical scenarios). All 800 scenarios generated by SBST are critical, i.e., FF1 $\leq$ 1 m or FF3 $\leq$ 0.5 s in all scenarios.

Among the 400 scenarios generated by PreScan and Pro-SiVIC, 229 (57.3\%) and 236 (59.0\%) scenarios led to safety violations, respectively. For the remaining 171 (42.8\%) scenarios of PreScan and the remaining 164 (41.0\%) scenarios of Pro-SiVIC, the pedestrian was detected by PeVi. However, in all those non-safety violation scenarios, the simulators still recorded collisions between the car and the pedestrian. Note that PeVi only provides a warning and does not apply any braking. More precisely, we found that 396 out of 400 (99.0\%) of the PreScan scenarios resulted in collisions between the car and the pedestrian. For Pro-SiVIC, the corresponding figure was 345 (86.3\%).
 
The right part of Figure~\ref{fig:table_hv} shows the distributions of the HV indicators computed based on the Pareto front outputs obtained from different runs of NSGA-II in PreScan and Pro-SiVIC, respectively. 
There is no statistically significant difference between the two HV distributions, indicating that the quality of the Pareto front outputs obtained from PreScan and Pro-SiVIC are comparable.

\begin{framed}
\noindent The principal findings from the original PreScan study can be reproduced using Pro-SiVIC. SBST is an effective approach to ADAS testing and the quality of the generated scenarios is comparable across simulators.
\end{framed}

\subsection{RQ2: X-sim Reproduction of Diagnosis Information} \label{sec:resrq2}
\begin{figure*}
\centering
\includegraphics[width=\textwidth]{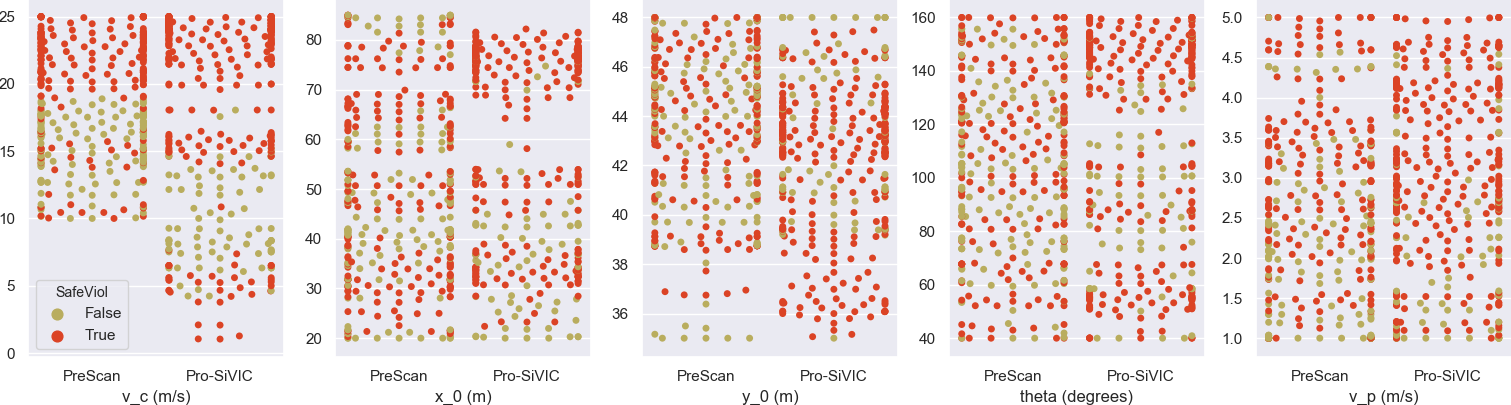}
\caption{Distribution of input parameters ($v_0^c, x^p_0, y^p_0, \theta^p$, and $v^p_0$) for the 800 test scenarios generated by applying SBST in PreScan and Pro-SiVIC. Red points denote safety violations (229 in PreScan and 236 in Pro-SiVIC).}
\label{fig:input_swarm}
\end{figure*}

Figure~\ref{fig:input_swarm} depicts swarm plots for the input parameters of the 800 test scenarios generated using SBST in PreScan and Pro-SiVIC, respectively. The red points represent safety violations (229 for PreScan and 236 for Pro-SiVIC). Note that box plots are not an appropriate visualization format, as some distributions are not only skewed but there are also considerable gaps in the data resulting in multimodal distributions. As Figure~\ref{fig:input_swarm} shows, SBST found effective test inputs (i.e., test inputs revealing critical behaviors of PeVi) in most areas of the input space, but there is a notable exception. Using PreScan, no effective test scenarios involved the car driving slower than 10 m/s ($v^c_0$), i.e., 36 km/h. Moreover, for both simulators, there are certain parameter ranges that are considerably sparser compared to the rest the input space, e.g., $55  \leq x_0\mbox{(m)} \leq 68$ in Pro-SiVIC and $y_0 \leq 38\mbox{(m)}$ in PreScan. Those ranges are not consistent between PreScan and Pro-SiVIC, illustrating internal variations when generating test scenarios using different simulators.

Further analysis of Figure~\ref{fig:input_swarm} shows that  diagnostic information obtained by PeVi testing are different in the two simulators. Specifically, in that figure, red points show test inputs resulting in safety violations. The distribution of red points show how the PeVi safety violations cluster for some parameter ranges -- but again, the results are not consistent between the two simulators. For example, Pro-SiVIC testing with $15 \leq v_c~(m/s) \leq 17$ result in many safety violations, for which the PreScan counterpart paints a different picture. Another divergent example is visible at $v_p \approx 4.4$ m/s, where Pro-SiVIC identifies safety violations but PreScan does not.

\begin{figure*}
\centering
\includegraphics[width=0.85\textwidth]{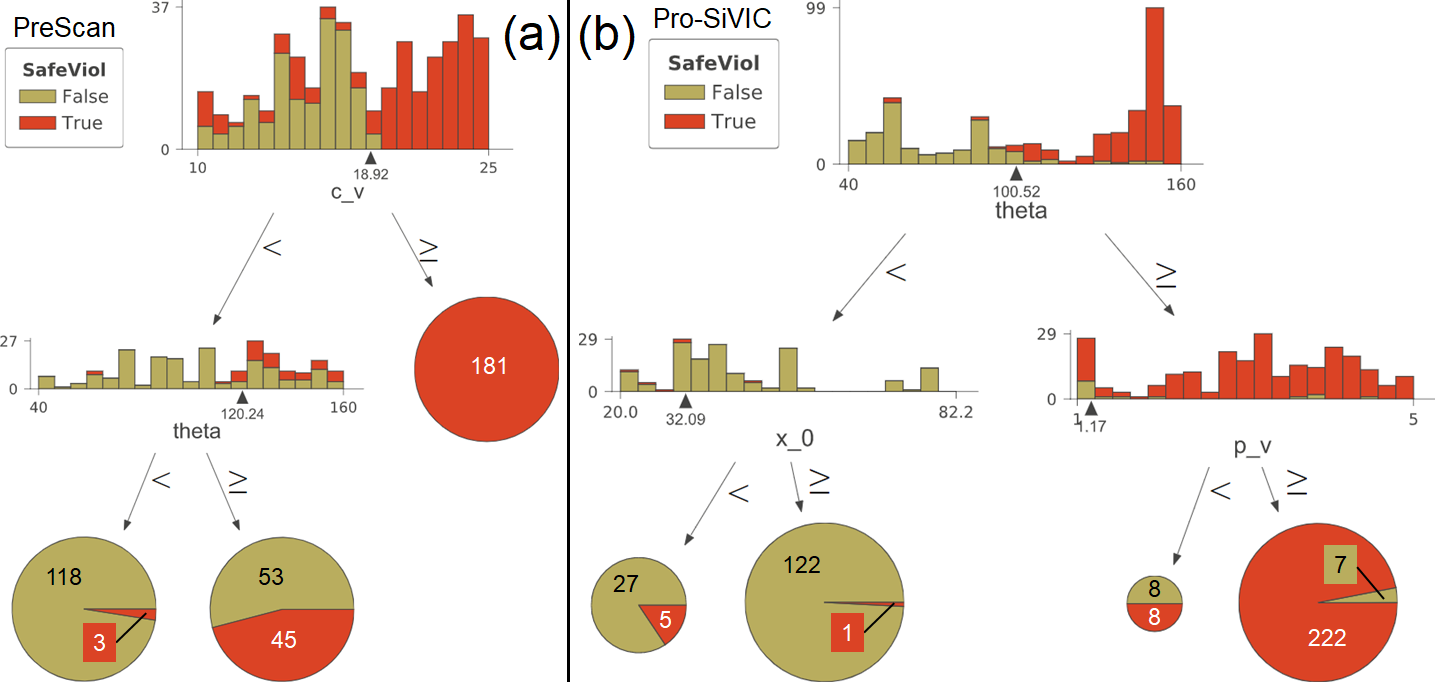}
\caption{Decision trees explaining when safety violations occur in PreScan (a) and Pro-SiVIC (b), respectively.}
\label{fig:explain_safeviol}
\end{figure*}



Figure~\ref{fig:explain_safeviol} displays two decision trees that we have built based on the 400 test inputs generated by PreScan (a) and Pro-SiVIC (b). 
The test inputs for both simulators are labelled by True (when they lead to a safety violation) and by False, otherwise. 
For example, Figures~\ref{fig:explain_safeviol}(a) shows that in 181 safety violation scenarios  generated by PreScan, the speed of the car was more than $18.92$ m/s (68 km/h), and in 45 other safety violation scenarios, the speed of the car was less than $18.92$ m/s (68 km/h) but the orientation of the pedestrian was more than $120.24^\circ$. The latter conditions, however, characterize a leaf of the tree with a mix of True and False-labelled output, and hence, cannot be taken as a  characterization of failures. Similarly, Figure~\ref{fig:explain_safeviol}(b) shows that for Pro-SiVIC the majority of safety violations are characterized by the conjunction of two conditions: $\theta$ ($\geq 100.52^\circ$) and $v^p \geq 1.17 m/s$. 

Overall, based on the test results in Figure~\ref{fig:input_swarm} and the decision trees in Figures~\ref{fig:explain_safeviol}(a) and (b), we can identify only one condition that can consistently  characterize safety violations identified by both Pro-SiVIC and PreScan. In particular, in both PreScan and Pro-SiVIC, PeVi performs worse when the car moves fast. More specifically, when $v_c \geq 20$ m/s (72 km/h), all test scenarios generated by PreScan and Pro-SiVIC lead to safety violations. However, apart from this condition, there are few patterns that can characterize safety violations for PeVi.

\begin{framed}
\noindent The results obtained by PreScan and Pro-SiVIC do not generally lead to consistent and conclusive characterizations of safety violations for PeVi. The only consistent conclusion is that PeVi likely violates its safety safety requirement when the car moves fast ($\geq$ 72 km/h).
\end{framed}

\subsection{RQ3: X-Sim Reproduction of Critical Test Scenarios} \label{sec:resrq3}
\begin{figure}
\centering
\includegraphics[width=0.48\textwidth]{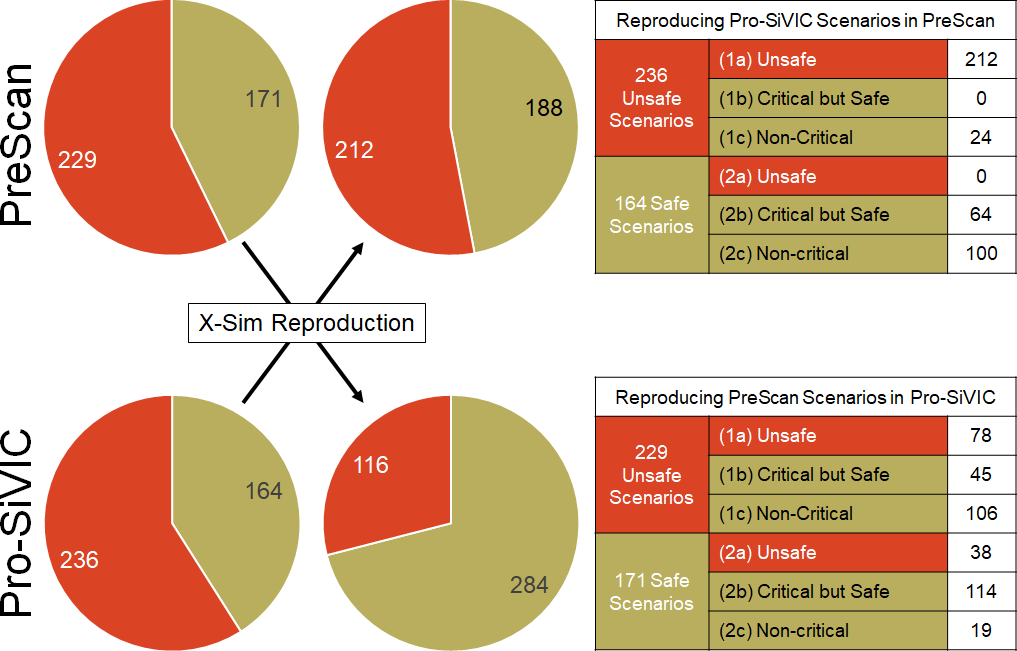}
\caption{Results from X-Sim reproduction of critical scenarios between PreScan and Pro-SiVIC.}
\label{fig:xsim-safety}
\end{figure}

Recall from Section~\ref{sec:resrq1} that all the 800 scenarios generated by PreScan and Pro-SiVIC were critical but some led to safety violations and some did not (see the definitions of critical and safety violations in Section~\ref{subsec:original}). To simplify the discussion, we refer to scenarios as either \textit{unsafe} (when they lead to a safety violation) or \textit{safe} otherwise. Figure~\ref{fig:xsim-safety} displays the results from X-sim reproduction of critical scenarios between PreScan and Pro-SiVIC. Before discussing the figure, we present the six possibilities that can happen  when executing a critical scenario generated by one simulator (SimA) in another simulator (SimB). The references in parentheses below refer to rows in Figure~\ref{fig:xsim-safety}.
 
\textbf{-} An unsafe scenario in SimA can: (1a) 
   also be unsafe in SimB (the detection failed in both SimA and SimB); (1b) be critical but become safe in SimB (the detection failed only in SimA); and (1c) be non-critical in SimB (in SimB, neither FF1 nor FF3 is small enough to warrant the scenario as critical).

\textbf{-} A safe scenario in SimA can: (2a) be unsafe in SimB (the detection failed only in SimB); (2b) be both critical and safe in SimB (the detection works in both SimA and SimB); and (2c) be non-critical in SimB (same reason as in 1c).

As shown in Figure~\ref{fig:xsim-safety}, after reproducing the scenarios generated by PreScan in Pro-SiVIC, we obtain the following results: Out of the 229 unsafe scenarios generated by PreScan, 78 (34.1\%) are unsafe (1a) and 151 are safe (65.9\%) (1b+1c) in Pro-SiVIC, but 45 of these 151 scenarios are still critical (1b) in Pro-SiVIC and the remaining 106 scenarios turn out to be non-critical and safe (1c) in Pro-SiVIC. Specifically, in the 45 scenarios (1b), the PeVi detection fails in PreScan but works in Pro-SiVIC, and in the 106 scenarios (1c), the distances between the ego car and the pedestrian in time and space in Pro-SiVIC are large enough to no longer constitute a critical scenario. 
 
Among the 171 safe but critical scenarios generated by PreScan, 133 (78.8\%) are safe (2b+2c), out of which 114 (2b) are still critical while 19 (2c) are no longer critical. The remaining 38 scenarios (2a) change from being safe (but critical) in PreScan to unsafe in Pro-SiVIC, indicating that PeVi detected the pedestrian in PreScan but failed in Pro-SiVIC. In short, the discrepancies in X-sim reproduction of critical scenarios from PreScan and Pro-SiVIC were due the following factors: (1) Inconsistencies in detecting the pedestrian (for 45 scenarios, the PeVi detection worked in Pro-SiVIC but failed in PreScan; and in 38 scenarios, the detection worked in PreScan but not in Pro-SiVIC); (2) Changes in the distances between the ego car and the pedestrian in time and space. Specifically, 125 scenarios (1c+2c) that were critical in PreScan turned out to be non-critical in Pro-SiVIC. 
 

When reproducing the 400 Pro-SiVIC scenarios in PreScan, among the 236 unsafe scenarios in Pro-SiVIC, 212 (89.8\%) are unsafe in PreScan (1a), and all the 164 safe scenarios in Pro-SiVIC are safe in PreScan as well (2b+2c). Among the 236 unsafe scenarios in Pro-SiVIC, 24 are safe (10.2\%) because the scenarios are no longer critical (1c). We observed no discrepancies in PeVi's pedestrian detection after X-sim reproduction from Pro-SiVIC to PreScan (1b+2a). However, we found that 100 of the 164 (61.0\%) safe scenarios became non-critical when reproducing in Pro-SiVIC (2c), indicating changes in the distances between the ego car and the pedestrian in time and space.

The X-sim reproductions show that the dynamic simulator models can substantially influence the test results. First, critical scenarios often turned non-critical as distances changed (1c+2c). Second, the PeVi detections are not necessarily consistent between the simulators, i.e., the overall test verdicts related to safety violations frequently differ when reproducing scenarios in Pro-SiVIC (1b+2a). This might also be due to the dynamic modeling of the pedestrian, but it can also be explained by the implementation of PeVi or differences in the off-the-shelf sensors available in the simulators' libraries.

\begin{figure*}
\centering
\includegraphics[width=0.99\textwidth]{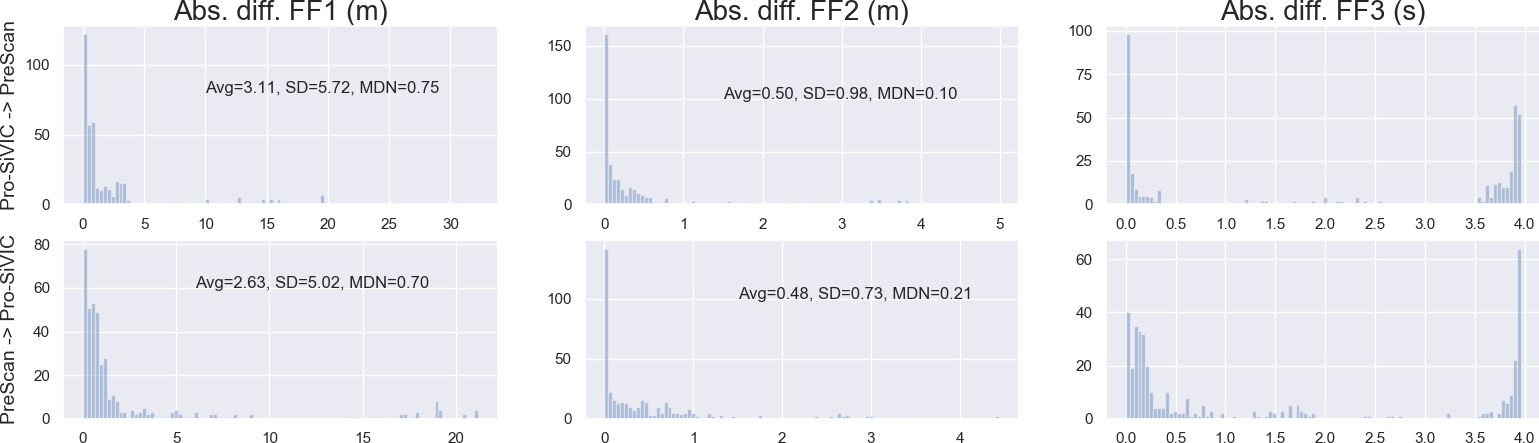}
\caption{Absolute differences of the three FFs when reproducing 400 critical Pro-SiVIC scenarios in PreScan (top row) and 400 critical PreScan scenarios in Pro-SiVIC (bottom row).}
\label{fig:diff_matrix}
\end{figure*}

To know which fitness function (among FF1, FF2, and FF3) contributes most to the discrepancies in the X-sim reproduction results of critical scenarios between PreScan and Pro-SiVIC, we measure how big the FF differences are after the X-sim reproductions. Figure~\ref{fig:diff_matrix} shows distributions of absolute differences for the three FFs. The top row shows the results from reproducing the 400 critical Pro-SiVIC scenarios in PreScan. For most scenarios, the difference in FF1 is less than 4 m, i.e., $FF1 \leq 4$ m. However, for 59 scenarios (14.8\%), the absolute difference in FF1 is $\geq$ 5 m. Furthermore, the distribution of differences for FF2 resembles FF1 with an absolute difference of $\leq 1 m$ in 348 scenarios (87.0\%). 

The absolute differences in FF3 display a bimodal distribution. Due to a PeVi implementation choice, an FF3 value of 4 s either means that the sensor did not detect the pedestrian or the pedestrian remained far away during the entire scenario. Consequently, an absolute difference close to 4 s for equivalent test scenarios in PreScan and Pro-SiVIC is the result of either (1) large variations in how close the pedestrian gets to the car (as shown for the absolute difference in FF1) or (2) conflicting results of the sensors in PreScan and Pro-SiVIC. Reproduction of 400 critical Pro-SiVIC scenarios in PreScan resulted in 156 scenarios with an absolute difference for FF3 of $\leq 0.5 s$ (41.3\%) and 195 scenarios of $\geq 3.5 s$ (48.8\%).

The bottom row in Figure~\ref{fig:diff_matrix} depicts the results from reproducing the 400 critical PreScan scenarios in Pro-SiVIC. The results largely resembles the Pro-SiVIC to PreScan reproduction. For FF1, the absolute difference is $\geq 4 m$ in 62 scenarios (15.5\%). For FF2, the absolute difference is $\geq 1 m$ in 48 scenarios (12.0\%). Finally, a bimodal distribution is again the result for FF3, the absolute difference is $\leq 0.5 s$ in 214 scenarios (53.5\%) and $\geq 3.5 s$ in 118 scenarios (29.5\%).

In short, FF2 values were the most consistent in the two simulators; FF1 values were largely consistent with a few outlier scenarios for which FF1 differences were large between PreScan and Pro-SiVIC; but FF3 values were the most inconsistent between the two simulators where we observed large FF3 differences in the results  from PreScan and Pro-SiVIC for several scenarios.

\begin{framed}
\noindent Reproducing critical scenarios between PreScan and Pro-SiVIC frequently results in discrepancies regarding distances and PeVi detections. Among the three fitness functions used for scenario generation, FF2 values were the most consistent after X-sim reproductions whereas FF3 values differed substantially.
\end{framed}

\section{Threats to Validity} \label{sec:threats}
In this section, we discuss the most important threats to internal and external validity~\cite{wohlin2012experimentation}.  

\textbf{Internal validity} concerns inferences regarding casual relationships. We designed our experiment setup in a way to mitigate internal  threats. For the X-sim reproductions, we focused on controlling as many variables as possible, both in the initial minimalistic scene and in the SBST setup. We used the same NSGA-II parameters X-sim and carefully created highly similar minimalistic scenes in PreScan and Pro-SiVIC. As described in Section~\ref{sec:porting}, there are some minor visual differences related to the initial scene. Furthermore, there are several other variables embedded in the simulators whose effects we study rather than control, e.g., the modeling of the optics in the cameras, sensor resolution and the radar cross-section of pedestrians. To mitigate this threat, we tried to ensure that sensors used in PreScan and Pro-SiVIC were configured the same to the extent possible. Finally, the packet loss measured in the Pro-SiVIC setup might have influenced the results, but as reported in Section~\ref{sec:designrq2}, we mitigated this by repeating the Pro-SiVIC experiments.


\textbf{External validity} reflects generalizability and often contradicts the internal validity of a study. As we carefully controlled variables through a minimalistic scene, we have no evidence regarding simulation of more complex environments or traffic scenarios. Future work should explore X-sim reproduction for more complex scenes, including variations in elevation and curvature of roads as well as scenery and traffic density. Another variable that deserves future study is how different weather simulations influence the sensor models. However, since we contribute evidence of prevalent X-sim discrepancies for a minimalistic scene, we have no reason to believe they would disappear for more complex scenes. Finally, as we limited the study to X-sim reproduction between PreScan and Pro-SiVIC, we cannot claim that the magnitude of differences would be the same for other simulators such as CARLA~\cite{dosovitskiy2017carla}.

\section{Lessons Learned} \label{sec:lessons}
Conducting X-sim reproductions between two commercial simulators provided insights that go beyond the RQs. In this section, we provide two important lessons learned.

\emph{Lesson~1. Validating ADAS testing results in multiple simulators is beneficial.} Automotive simulators are complex software tools. Each simulator depends on the priorities, background and expertise of its vendor and is focused on certain aspects of ADAS and self-driving systems. For example, while PreScan is mostly focused on the fidelity of physics-based and mechanical models of self-driving systems, Pro-SiVIC is specialized in developing accurate sensor models. Typically, one simulator alone may not be able to perfectly capture all the subtleties and complexities of ADAS and self-driving systems. As a result, replicating simulations in multiple tools with complementary strengths, identifying simulations that are consistent and robust across tools, and using those simulations for failure analysis and fault localization can help improve the accuracy of ADAS V\&V activities. If the ADAS performance cannot generalize to another simulator, it would be overly optimistic to expect generalization to the real world.

\emph{Lesson~2. Fitness functions (test objectives) should be defined in such a way that they are minimally impacted by variations, weaknesses or potential faults in the internals of simulators.} In our replication study, we observed different results for the three fitness functions that were proposed in the original study. Specifically, FF2 values were the most consistent in the two simulators and FF1 values were largely consistent with a few outlier scenarios; but FF3 values were the most inconsistent between the two simulators. While FF1 and FF2 largely depend on the physics-based models of simulators, FF3 depends on their sensor models and could, at least in part, explain the measured differences. Our observation is that testing results might be more consistent across different simulators if fitness functions do not depend on specific and non-standard components (e.g., sensors) that likely vary across different simulators. For example, we conjecture that if we repeat our  study where we use the same SBST method with FF1 and FF2 only and remove FF3, we likely obtain more consistent test outputs between PreScan and Pro-SiVIC.


\section{Conclusion and Future Work} \label{sec:conc}
We presented a replication study of applying a search-based software testing (SBST) solution to an Advanced Driver-Assistance System (ADAS) case study using two different commercial simulators, namely, TASS/Siemens PreScan and ESI Pro-SiVIC. Our results suggest that while SBST is effective in finding failure revealing test scenarios using both simulators, there are considerable differences in the specific details of the scenarios generated using the two simulators. 

We present two recommendations for research and practice. First, simulation-based ADAS testing should not rely on a single simulator. Ideally, the test result analysis should primarily be based on the ADAS testing results that generalize to multiple simulators. Second, SBST for ADAS testing should be based on fitness functions (test objectives) that are minimally impacted by the internals of simulators and in particular by third party models of hardware components (e.g., sensors and cameras) included in the simulators.

For future, we will elaborate on how our findings relate to the evaluation of the residual risk as mandated in ISO/PAS 21448 Safety of the Intended Function (SOTIF). Specifically, we will provide actionable recommendations for the standardization efforts related to SOTIF Part 11.2 -- Method K ``Simulation of selected scenarios''. As our primary interests relate to testing of SOTIF compliant perception systems that use deep neural networks, our next step will be development of testing recommendations tailored for machine learning.

\section*{Acknowledgment}
This work was carried out within the SMILE~II and SMILE~III projects financed by Vinnova, FFI, Fordonsstrategisk forskning och innovation under the grant numbers:  2017-03066 and 2019-05871. Further, the work has received funding from the ECSEL Joint Undertaking (JU) under grant agreement No 876852 (VALU3S), the European Research Council (ERC) under the European Union's Horizon 2020 research and innovation programme (grant agreement No 694277), and the NSERC of Canada under the Discovery program.
Donghwan Shin was partially supported by Basic Science Research Program through the National Research Foundation of Korea (NRF) funded by the Ministry of Education (2019R1A6A3A03033444).
\bibliographystyle{IEEEtran}
\bibliography{IEEEabrv, sim_refs}

\end{document}